\documentclass[runningheads]{llncs}
\usepackage{graphicx}
\usepackage{float}
\usepackage{todonotes}
\usepackage{tablefootnote}
\begin{document}
\title{Prediction of mandibular ORN incidence from 3D radiation dose distribution maps using deep learning}
\titlerunning{Prediction of ORN from 3D radiation dose distribution maps}
%

\author{Laia Humbert-Vidan\inst{1,2} \and
Vinod Patel\inst{3} \and
Robin Andlauer\inst{4} \and
Andrew P King\inst{4} ** \and
Teresa Guerrero Urbano\inst{5} \thanks{Joint last authors.} }

\authorrunning{L. Humbert-Vidan et al.}
%
\institute{Department of Medical Physics, Guy's and St Thomas' NHS Foundation Trust, London, UK\\
\and
School of Cancer and Pharmaceutical Sciences, Comprehensive Cancer Centre, King’s College London, London, UK\\
\email{laia.humbert-vidan@kcl.ac.uk}\\ 
\and
Department of Oral Surgery, Guy's Dental Hospital, London, UK
\email{Vinod.Patel@gstt.nhs.uk}\\
\and
School of Biomedical Engineering \& Imaging Sciences, King's College London, London, UK\\
\email{robin.andlauer@web.de}\\
\email{andrew.king@kcl.ac.uk}\\
\and
Department of Clinical Oncology, Guy's and St Thomas' NHS Foundation Trust, London, UK\\
\email{Teresa.GuerreroUrbano@gstt.nhs.uk}
}
\maketitle              
\begin{abstract}

\emph{Background}. Absorbed radiation dose to the mandible is an important risk factor in the development of mandibular osteoradionecrosis (ORN) in head and neck cancer (HNC) patients treated with radiotherapy (RT).  
The prediction of mandibular ORN may not only guide the RT treatment planning optimisation process but also identify which patients would benefit from a closer follow-up post-RT for an early diagnosis and intervention of ORN.
Existing mandibular ORN prediction models are based on dose-volume histogram (DVH) metrics that omit the spatial localisation and dose gradient and direction information provided by the clinical mandible radiation dose distribution maps.
\emph{Methods}. We propose the use of a binary classification 3D DenseNet121 to extract the relevant dosimetric information directly from the 3D mandible radiation dose distribution maps and predict the incidence of ORN. We compare the results to a Random Forest ensemble with DVH-based parameters. 
\emph{Results}.
The 3D DenseNet121 model was able to discriminate ORN vs. non-ORN cases with an average AUC of 0.71 (0.64-0.79), compared to 0.65 (0.57-0.73) for the RF model.
\emph{Conclusion}.   Obtaining the dosimetric information directly from the clinical radiation dose distribution maps may enhance the performance and functionality of
ORN normal tissue complication probability (NTCP) models.


\end{abstract}

\section{Introduction}

Radiotherapy (RT), either alone or combined with surgery and/or chemotherapy, is typically the primary treatment for head and neck cancer (HNC). With the introduction of modern RT techniques, we are able to irradiate the target volume with high precision and dose conformity. However, due to the nature of energy deposition of photons in tissue, non-target organs inevitably absorb ionising radiation, potentially resulting in normal tissue toxicity. Acute and late radiation-induced toxicities have a significant effect on the patient's quality of life and can also jeopardise treatment compliance, potentially impacting treatment outcome.

Osteoradionecrosis (ORN) of the mandible is a rare but severe radiation-induced toxicity observed in 4-8\% \cite{Frankart2021} of patients after HNC RT. Radiation damages the vascularisation of the mandible. Consequently, necrosis of the bone can develop either spontaneously or be triggered by trauma to the mandible bone (e.g. dental extractions, surgery, implants). Necrosis occurs because the bone is not able to heal due to reduced blood supply, hypoxia or hypo-cellularity caused by exposure to radiation \cite{Chen2016}. The severity of ORN can vary between patients with the most severe cases experiencing significant pain levels and even pathological fracture of the mandible (Figure 1).

\begin{figure}[htbp]
\centering
\includegraphics[width=0.4\linewidth]{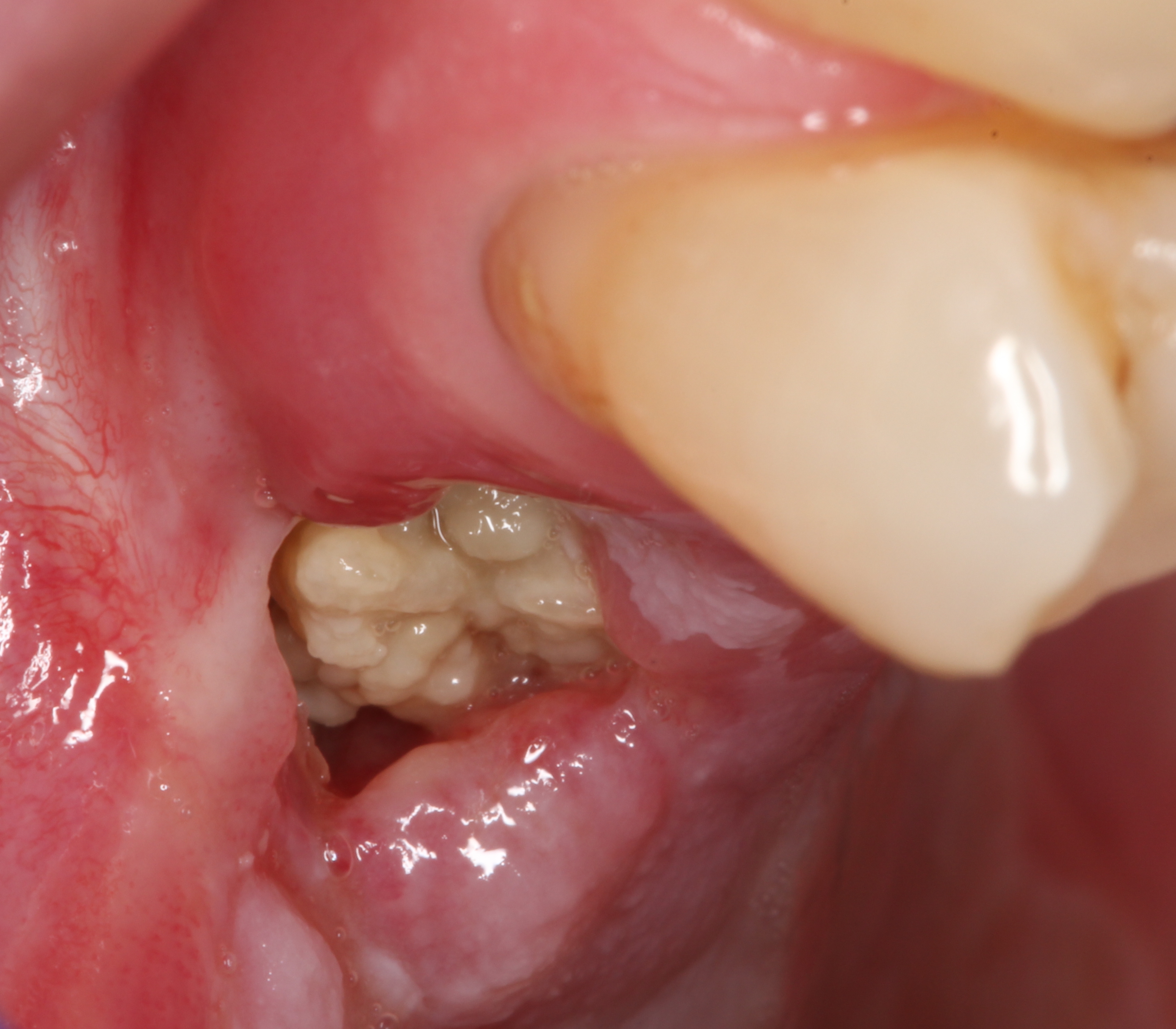}
\caption{Example of a recent mandibular ORN case at our centre. This image shows exposed bone in line with ORN of the upper right second premolar tooth socket.} \label{fig1}
\end{figure}

Quality of life can be greatly affected in patients who develop mandibular ORN. Moreover, the management of ORN is often complex and requires costly clinical interventions \cite{Patel2017}. Being able to identify patients that are at a higher risk of developing ORN will contribute to more individualised treatment and follow-up regimes potentially resulting in a reduced incidence probability or early management with improved prognosis.

The amount of an organ irradiated to a certain dose level is typically described in the clinical setting using a 2D dose-volume histogram (DVH), where the horizontal axis is divided into dose bins and the vertical axis represents the percentage or absolute volume of the organ receiving each of the dose levels on the horizontal axis. Thus, in a DVH the 3D radiation dose distribution map of an organ is reduced to a 2D representation with the resulting loss of clinically relevant spatial localisation as well as dose gradient and direction information. Spatial dosimetric information is of great interest in the investigation of radiation damage to the mandible, where there are regions that are more vulnerable to ORN development such as the posterior molar segment \cite{Habib2021}. In addition, it has previously been observed \cite{HumbertVidan2021a} that the ORN region within the mandible may actually develop far from the high dose region. 

Normal tissue complication probability (NTCP) models are used as a clinical decision support system to reduce the incidence of a given toxicity by identifying the patients who are at a higher risk of developing it. Existing ORN NTCP models are based on DVH data \cite{VanDijk2021,HumbertVidan2021b}. The use of spatial dose features in NTCP models may provide more comprehensive information than the typically used DVH-based metrics such as mean or maximum doses. Other HNC studies \cite{Beasley2018,Gabrys2018,Dean2018} have included spatial dose metrics into NTCP models by manually extracting dosiomic features on a voxel-by-voxel basis. An alternative method for incorporating spatial dose metrics into NTCP models is the use of deep learning (DL) \cite{Ibragimov2018,Men2019}.

To the best of our knowledge, this is the first study to investigate the use of DL methods in mandibular ORN NTCP models with clinical 3D radiation dose distribution maps as input. 

\section{Methods and Materials}

\subsection{Data}

\subsubsection{Patient Selection:}
A cohort of 70 ORN cases and 70 control cases was retrospectively selected from a database of HNC cases treated at our hospitals with radical intensity-modulated RT (IMRT) between 2011 and 2019. Table 1 provides the demographic and clinical characteristics and compares the ORN and control cohorts.
During the time span considered, a total 187 patients were diagnosed with ORN. From the entire ORN population, 36 cases were treated with 3D conformal RT instead of IMRT, 9 cases were not treated for HNC, for 32 cases the RT dose and/or RT plan DICOM files were unavailable and 40 additional patients were excluded either due to previous irradiation of the HN region, the ORN being located in the maxilla or receiving a palliative treatment.

At our institution ORN is graded according to its severity following the Notani grading system, which classifies ORN severity into three categories \cite{Notani2002}. However, for the purpose of binary classification in this study, any grade of ORN was considered as an event.

\begin{table}
\centering
\caption{Demographic and clinical characteristics.}\label{tab1}
\begin{tabular}{|l|c|c|c|}
\hline
\bfseries & \bfseries ORN & \bfseries Control & \bfseries p-value \\
\hline
Gender & & & \\
\emph{Male/Female} (\%) & 70.0/30.0 & 80.0/20.0 & 0.15/0.15  \\
\hline
Age (median (IQR))& 61.5 (13.3) & 60.5 (13.5) & 0.49 \\
\hline
Primary tumour site & & & \\
\emph{Oropharynx} (\%) & 60.0 & 52.1 & 0.35 \\
\emph{Oral cavity} (\%) & 30.0 & 21.4 & 0.23 \\
\emph{Larynx} (\%) & 2.9 & 12.9 & 0.04 \\
\emph{Hypopharynx} (\%) & 0.0 & 2.9 & 0.37 \\
\emph{Salivary glands} (\%) & 1.4 & 3.6 & 0.66 \\
\emph{Nasopharynx} (\%) & 0.0 & 1.4 & 0.80 \\
\emph{Paranasal sinus} (\%) & 1.4 & 0.0 & 0.72 \\
\emph{Unknown primary} (\%) & 4.3 & 2.1 & 0.66 \\
\hline
Smoking & & & \\
\emph{Current/Previous} (\%) & 44.3/27.1 & 32.1/36.4 & 0.12/0.23  \\
\hline
Alcohol & & & \\
\emph{Current/Previous} (\%) & 62.9/10.0 & 63.6/9.3 & 0.96/0.93 \\
\hline
Chemotherapy (\%) & 65.7 & 62.1 & 0.72 \\
\hline
Pre-RT dental extractions (\%) & 64.3 & 63.6 & 0.96 \\
\hline
Pre-RT surgery (\%) & 32.9 & 32.1 & 0.96 \\
\hline

\hline
\end{tabular}
\end{table}

\subsubsection{3D Mandible Dose Distribution Maps:}
The data preparation workflow is illustrated in Figure 2. The mandible was manually segmented from the patients' CT volumes by a single observer in the treatment planning system (TPS) including the mandible sockets and excluding the maxilla and teeth. The 3D radiation dose maps, mandible segmentations and CT images were exported as DICOM files from the RT TPS. All data were resampled to a common slice thickness of 2 mm and slice size of 512 pixels x 512 pixels using 3-D Slicer. All mandible segmentations were rigidly registered to a common reference space using ITKSnap in order to reduce inter-patient positional variation. The patient with the largest number of mandible segmentation slices was selected as the reference patient and smaller mandibles were padded with empty slices. The dose maps were transformed using the same rigid transformations to maintain alignment. The 3D mandible dose distribution maps used by the model were obtained by multiplying the normalised dose maps by the binary mandible segmentation masks. Finally, the 3D mandible dose distribution maps were normalised to the voxel value range of the entire dataset.

\subsubsection{DVH metrics:}
The cumulative DVH of the mandible structure was exported in relative volume and absolute dose (Gy) from the TPS. Maximum (Dmax and D2\%), minimum (Dmin and D98\%), Dmean and Dmedian (D50\%) dose metrics were extracted from the DVH using the DVHmetrics package in R statistical software (R Foundation for Statistical Computing, Vienna, Austria). Dose-volume data were converted to an equivalent dose in 2 Gy fractions assuming an alpha-beta ratio of 3 for late effects \cite{Williams1985}.

\subsection{Prediction Models}
A 3D densely-connected 121-layer (DenseNet121) \cite{Huang2017} convolutional neural network (CNN)  was implemented with the MONAI (\url{https://monai.io/}) Pytorch-based framework  for the purpose of binary classification of ORN vs. control cases. A softmax activation layer was added at the end of the network to obtain the predicted probability for each class. The categorical cross entropy objective loss function was used. The 3D DenseNet121 (blue branch in Figure 2) was trained on the 3D mandible dose distribution maps. Small 3D random rotation (-0.1 to 0.1 rad) and zoom (0.8 to 1.2) augmentations were applied to the training images. 
The prediction performance of the 3D DenseNet121 was compared to a Random Forest (RF) trained with DVH metrics only (maximum, minimum, mean and median doses).

\subsection{Model Evaluation}
In both models, a stratified nested cross-validation (CV) approach with ensemble learning was followed (Figure 3). The data were split into training, validation and test sets following a nested CV approach. In the outer loop of this procedure, a stratified 5-fold CV was applied by randomly splitting the data into training (80\%) and test (20\%) sets repeatedly. Hyperparameter optimisation was performed using a stratified 5-fold inner CV on the training set of the outer CV folds.
For each of the outer folds, the entire training set was used for training using the optimised hyperparameters and the prediction accuracy calculated on the held-out fold. 
For the final training, an ensemble of models was trained to improve generalisation performance and to reduce the sensitivity of the model performance to stochastic noise of the training. In this study, our ensemble model was created by randomly initialising each model five times and each time, training the model on the training set of the outer fold. Due to the stochastic randomness of the weight initialisation and the selection of mini-batches during training, this created five slightly different models for each outer fold. To calculate the prediction of this ensemble model, the predicted softmax probabilities of each of the five individual models were averaged for each class (i.e. soft voting). 

\subsection{Statistical Analysis}
The predictive performance of the models was assessed in terms of their discriminative ability  using the area under the receiver operating characteristic curve (ROC AUC). The DeLong nonparametric statistical test \cite{DeLong1988} was performed to compare the AUC of the two models using the pROC package \cite{pROC} with the statistical software R (\url{https://www.R-project.org/}).

\begin{figure}[htbp]
\centering
\includegraphics[width=0.9\linewidth]{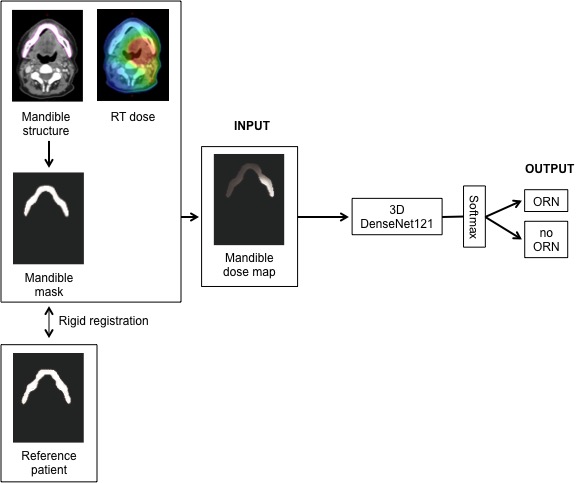}
\caption{Schematics of the data preparation workflow and deep-learning pipeline used} \label{fig2}
\end{figure}

\begin{figure}[htbp]
\centering
\includegraphics[width=0.8\linewidth]{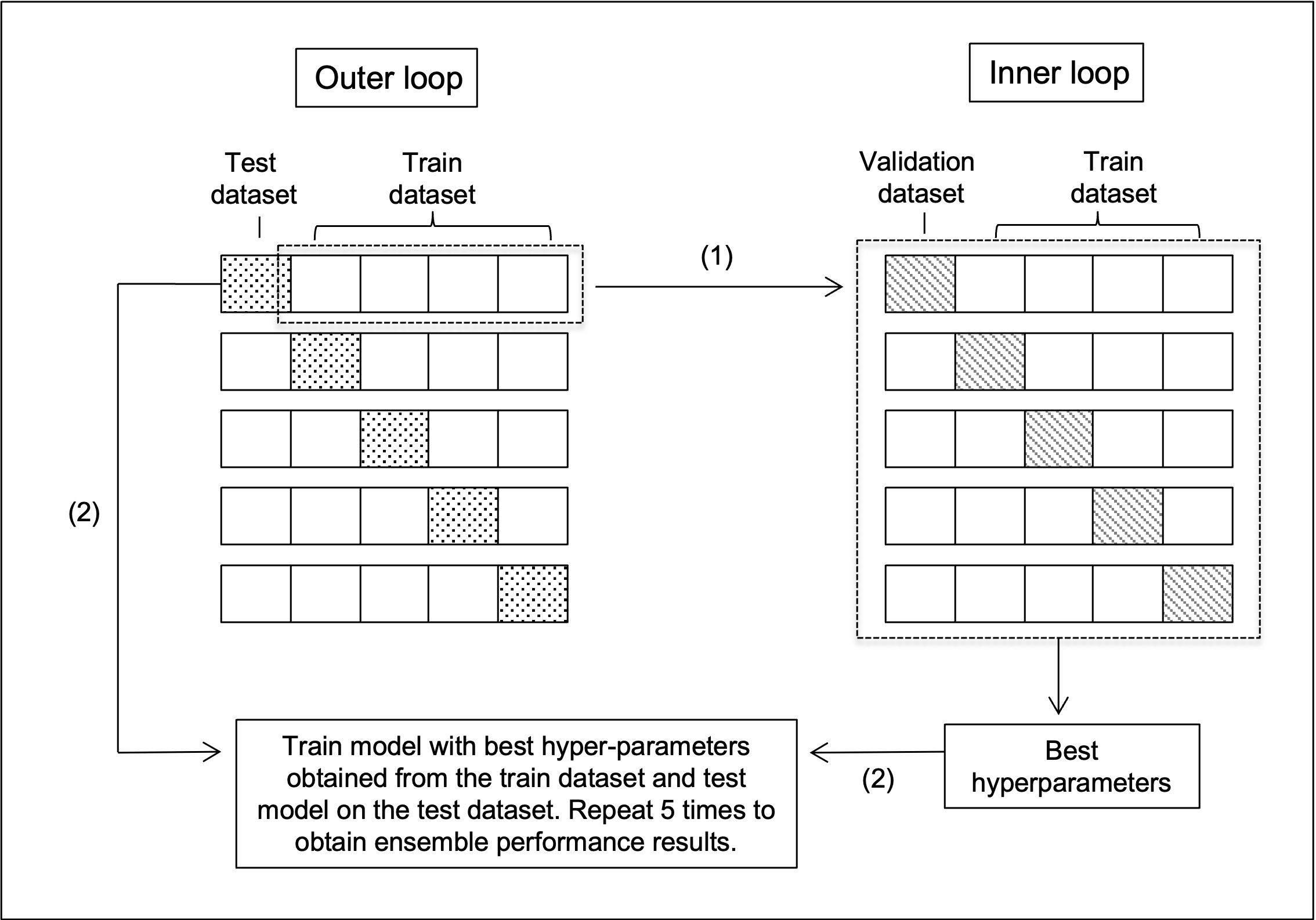}
\caption{Model evaluation workflow. A stratified nested CV approach with ensemble training was followed.} \label{fig3}
\end{figure}

\section{Results}
Table 2 summarises the model discrimination performance results obtained from the two models: a 3D DenseNet121 model trained with mandible dose distribution maps and a RF model trained with DVH metrics.
When comparing the two ROC curves with the DeLong test, the difference in AUC (0.71 vs. 0.65) was not found to be statistically significant with a p-value of 0.24 (significance level of 0.05).  

\begin{table}
\centering
\caption{Model discrimination performance.}\label{tab2}
\setlength{\tabcolsep}{10pt}
\begin{tabular}{|l|c|c|}
\hline
\bfseries & \bfseries 3D DenseNet121 & \bfseries Random Forest \\

 & (dose map) & (DVH metrics) \\
\hline
ROC AUC (95\% CI) & 0.71 (0.64-0.79) & 0.65 (0.57-0.73) \\
Sensitivity & 0.70 & 0.66 \\
Specificity & 0.73 & 0.64 \\
Precision & 0.72 & 0.65  \\
\hline
\end{tabular}
\end{table}

\section{Discussion}
In this study, we have explored the use of DL models to predict mandibular ORN in HNC using clinical 3D radiation dose distribution maps. This is a novel approach to NTCP modeling for mandibular ORN as it uses the actual RT dose distribution rather than the more traditionally used DVH parameters. 

\subsection{ORN Prediction}
Our results show that the 3D DenseNet121 model was able to discriminate well between ORN and control cases based on the 3D mandible dose distribution maps. However, we found that the performance of the model was dependent on the test-train data split (i.e. classification performance varied between the outer loop CV folds). This may be due to the high variability in the anatomical localisation of the radiation dose distribution of our cohort, suggesting that training using a larger cohort will lead to improved classification accuracy and robustness. Moreover, in some of the outer loop CV folds, there was a large variation between the ensemble models, i.e. there was stochastic noise in the model training process for a given test-train data split. 

Although the difference in model performance was not found to be statistically significant, the average discriminative performance of the RF model using DVH data was 9\% lower than that of the 3D DenseNet121. This could be due to the fact that the entire DVH was not included in this model, i.e. only maximum, minimum, mean and median doses were included as variables, although we note that these are the variables typically used in clinical NTCP models. However, it is likely that the inclusion of the spatial information in the dose maps may contribute to an improved performance. Furthermore, there are features such as the mandible volume that can be extracted from the mandible dose distribution maps but are not DVH dosimetric parameters that have previously been associated with ORN incidence \cite{HumbertVidan2021a,Patel2020}. 

\subsection{Study Limitations and Future Work}
Due to mandibular ORN being a rare toxicity, the case numbers are naturally low. Although we have tried to mitigate this with data augmentation in the DL model, a larger ORN population would enable us to more thoroughly evaluate the potential of CNNs in ORN prediction. 

Radiation dose, in particular maximum dose, has typically been associated with ORN incidence. There are, however, other risk factors for mandibular ORN to be considered \cite{Aarup-Kirstensen2019,MDAnderson2017}. In this study we have focused on radiation dose as the only risk factor but the inclusion of non-dosimetric clinical parameters into the model would be of great clinical value. Moreover, there are cases where ORN develops away from the high radiation dose region within the mandible and the correlation between ORN incidence and intermediate or high radiation doses is less obvious. Particularly in such cases, non-dosimetric parameters may play an important role in the development of ORN. We will investigate including such clinical parameters in our CNN framework in future work.

Finally, in this study we have not included information on ORN region localisation. To take full advantage of a DL model using dose distribution maps, knowledge of the actual ORN region or at least the ORN localisation within the mandible may be included into the model. Predicted ORN incidence could then be analysed, for instance, with regards to proximity of the ORN localisation to the tumour volume or the high radiation dose region within the mandible.

\section{Conclusion}
This study is, to the best of our knowledge, the first to investigate the use of a deep CNN in mandibular ORN NTCP modeling. Our results suggest that it is not only possible to predict mandibular ORN incidence from 3D radiation dose distribution maps but that a deep CNN model may even outperform the more traditional DVH-based models. 

With the methods proposed in this study, we are contributing towards a more individualised treatment of HNC by predicting which patients are more likely to develop mandibular ORN based on their planned radiation dose distribution. This will not only guide the RT treatment planning optimisation process but also identify which patients would benefit from a closer follow-up post-RT for an early diagnosis and intervention of ORN.  Obtaining the dosimetric information directly from the clinical radiation dose distribution maps rather than the DVH may enhance the performance and functionality of the ORN NTCP models.



\begin{thebibliography}{8}
\bibitem{Frankart2021}
Frankart, AJ., Frankart, MJ., Cervenka, B., Tang, AL., Krishnan, DG and Takiar, V.: Osteoradionecrosis: Exposing the evidence not the bone. Int J Radiation Oncol Biol Phys (2021)

\bibitem{Chen2016}
Chen, JA., Wang, CC., Wong, YK., Wang, CP., Jiang, RS., Lin, JC., Chen, CC and Liu, SA.: Osteoradionecrosis of mandible bone in patients with oral cancer--associated factors and treatment outcomes. Head \& neck \textbf{38}, 762-768 (2016)

\bibitem{Patel2017}
Patel, V., Ormondroyd, L., Lyons, A. and McGurk, M.: The financial burden for the surgical management of osteoradionecrosis. British Dental Journal \textbf{222}, 177-180 (2017)

\bibitem{Habib2021}
Habib, S., Sassoon, I., Thompson, I. and Patel, V.: Risk factors associated with osteoradionecrosis. Oral Surgery \textbf{14}, 227-35 (2021)

\bibitem{HumbertVidan2021a}
Humbert-Vidan, L., Patel, V., Begum, RH., McGovern, M., Eaton, D., Kong, A., Petkar, I., Reis Ferreira, M., Lei, M and King, AP. and Guerrero Urbano, T.: PH-0387 Mandible osteoradionecrosis: a dosimetric study (poster presented at ESTRO 2021, Madrid, Spain). Radiotherapy and Oncology \textbf{161} (2021)

\bibitem{VanDijk2021}
Van Dijk, LV., Abusaif, AA., Rigert, J., Naser, MA., Hutcheson, KA., Lai, SY., Fuller, CD. and Mohamed, ASR.: Normal tissue complication probability (NTCP) prediction model for osteoradionecrosis of the mandible in patients with head and neck cancer after radiation therapy: large-scale observational cohort. Int J Radiation Oncol Biol Phys \textbf{111} (2), 549-558 (2021)

\bibitem{HumbertVidan2021b}
Humbert-Vidan, L., Patel, V., Oksuz, I., King, AP. and Guerrero Urbano, T.: Comparison of machine learning methods for prediction of osteoradionecrosis incidence in patients with head and neck cancer. Br J Radiol \textbf{94} (2021)

\bibitem{Beasley2018}
Beasley, W., Thor, M., McWilliam, A., Green, A., Mackay, R., Slevin, N., Olsson, C., Pettersson, N., Finizia, C., Estilo, C., Riaz, N., Lee, NY., Deasy, JO. and van Herk, M.: Image-based Data Mining to Probe Dosimetric Correlates of Radiation-induced Trismus. International journal of radiation oncology, biology, physics \textbf{102} (4), 1330-1338 (2018)

\bibitem{Jiang2019}
Jiang, W., Lakshminarayanan, P., Hui, X., Han, P., Cheng, Z., Bowers, M., Shpitser, I., Siddiqui, S., Taylor, RH., Quon, H. and McNutt, T.: Machine Learning Methods Uncover Radiomorphologic Dose Patterns in Salivary Glands that Predict Xerostomia in Patients with Head and Neck Cancer. Advances in radiation oncology \textbf{4} (2), 401-412 (2019)

\bibitem{Gabrys2018}
Gabryś, HS., Buettner, F., Sterzing, F., Hauswald, H. and Bangert, M. Design and selection of machine learning methods using radiomics and dosiomics for normal tissue complication probability modeling of xerostomia. Front Oncol \textbf{8} (25) (2018)

\bibitem{Dean2018}
Dean, J., Wong, K., Gay, H., Welsh, L., Jones, AB., Schick, U., Oh, JH., Apte, A., Newbold, K., Bhide, S., Harrington, K., Deasy, J., Nutting, C. and Gulliford, S.: Incorporating spatial dose metrics in machine learning-based normal tissue complication probability (NTCP) models of severe acute dysphagia resulting from head and neck radiotherapy. Clin Trans Radiat Oncol \textbf{8}, 27-39 (2018)

\bibitem{Ibragimov2018}
Ibragimov, B., Toesca, D., Chang, D., Yuan, Y., Koong, A. and Xing, L.: Development of deep neural network for individualized hepatobiliary toxicity prediction after liver SBRT. Medical physics (2018).

\bibitem{Men2019}
Men, K.,  Geng, H.,  Zhong, H., Fan, Y., Lin, A. and Xiao, Y.: A deep learning model for predicting xerostomia due to radiotherapy for head-and-neck squamous cell carcinoma in the RTOG 0522 clinical trial. Int J Radiat Oncol Biol Phys \textbf{105} (2), 440-447 (2019).

\bibitem{Notani2002}
Notani, K., Yamazaki, Y., Kitada, H., Sakakibara, N., Fukuda, H., Omori, K. and Nakamura, M.: Osteoradionecrosis of the mandible—factors influencing severity. Asian Journal of Oral Maxillofacial Surgery (2002)

\bibitem{Williams1985}
Williams, MV., Denekamps, J. and Fowler, JF.: A review of alpha/beta ratios for experimental tumours: implications for clinical studies of altered fractionation. Int J Radiat Oncol Biol Phys \textbf{11}, 87-96 (1985)

\bibitem{Huang2017}
Huang, G., Liu, Z., Van Der Maaten, L., and Weinberger, KQ.: Densely connected convolutional networks. Proceedings of the IEEE conference on computer vision and pattern recognition, 4700-4708 (2017)

\bibitem{DeLong1988}
DeLong, ER., DeLong D.M. and Clarke-Pearson D.L.: Comparing the Areas under Two or More Correlated Receiver Operating Characteristic Curves: A Nonparametric Approach. Biometrics \textbf{44}, 837-845 (1988)

\bibitem{pROC}
Robin, X., Turck, N., Hainard, A., Tiberti, N., Lisacek, F., Sanchez, J.C. and Müller, M.: pROC: an open-source package for R and S+ to analyze and compare ROC curves. BMC Bioinformatics \textbf{12}, 77 (2011)

\bibitem{Patel2020}
Patel, V., Humbert-Vidan, L., Thomas, C., Sassoon, I., McGurk, M., Fenlon, M. and Guerrero Urbano, T.: Radiotherapy quadrant doses in oropharyngeal cancer treated with intensity modulated radiotherapy. Faculty Dental Journal \textbf{11}, 166-72 (2015)

\bibitem{Aarup-Kirstensen2019}
Aarup-Kirstensen, S., Hansen, CR., Forner, L., Brink, C., Eriksen, JG. and Johansen, J.: Osteoradionecrosis of the mandible after radiotherapy for head and neck cancer: risk factors and dose-volme correlations. Acta Oncologica \textbf{58} (10), 1373-1377 (2019)

\bibitem{MDAnderson2017}
MDA HNC Symptom Working Group. Dose-volume correlates of mandibular osteoradionecrosis in oropharynx cancer patients receiving intensity-modulated radiotherapy: Results from a case-matched comparison. Radiother Oncol \textbf{124}, 232-239 (2017)

\end{thebibliography}
\end{document}